\documentstyle[12pt]{article}

\def\hybrid{\topmargin 0pt      \oddsidemargin 0pt
        \headheight 0pt \headsep 0pt
        \voffset=-0.5cm
        \textwidth 6.5in        % US paper
        \textheight 9in         % US paper
        \marginparwidth 0.0in
        \parskip 5pt plus 1pt   \jot = 1.5ex}
\catcode`\@=11
\def\marginnote#1{}

\newcount\hour
\newcount\minute
\newtoks\amorpm
\hour=\time\divide\hour by60
\minute=\time{\multiply\hour by60 \global\advance\minute by-\hour}
\edef\standardtime{{\ifnum\hour<12 \global\amorpm={am}%
        \else\global\amorpm={pm}\advance\hour by-12 \fi
        \ifnum\hour=0 \hour=12 \fi
        \number\hour:\ifnum\minute<10 0\fi\number\minute\the\amorpm}}
\edef\militarytime{\number\hour:\ifnum\minute<10 0\fi\number\minute}

\def\draftlabel#1{{\@bsphack\if@filesw {\let\thepage\relax
   \xdef\@gtempa{\write\@auxout{\string
      \newlabel{#1}{{\@currentlabel}{\thepage}}}}}\@gtempa
   \if@nobreak \ifvmode\nobreak\fi\fi\fi\@esphack}
        \gdef\@eqnlabel{#1}}
\def\@eqnlabel{}
\def\@vacuum{}
\def\draftmarginnote#1{\marginpar{\raggedright\scriptsize\tt#1}}

\def\draftlabel#1{{\@bsphack\if@filesw {\let\thepage\relax
   \xdef\@gtempa{\write\@auxout{\string
      \newlabel{#1}{{\@currentlabel}{\thepage}}}}}\@gtempa
   \if@nobreak \ifvmode\nobreak\fi\fi\fi\@esphack}
        \gdef\@eqnlabel{#1}}
\def\@eqnlabel{}
\def\@vacuum{}
\def\draftmarginnote#1{\marginpar{\raggedright\scriptsize\tt#1}}

\def\draft{\oddsidemargin -.5truein
        \def\@oddfoot{\sl preliminary draft \hfil
        \rm\thepage\hfil\sl\today\quad\militarytime}
        \let\@evenfoot\@oddfoot \overfullrule 3pt
        \let\label=\draftlabel
        \let\marginnote=\draftmarginnote
   \def\@eqnnum{(\theequation)\rlap{\kern\marginparsep\tt\@eqnlabel}%
\global\let\@eqnlabel\@vacuum}  }

\def\numberbysection{\@addtoreset{equation}{section}
        \def\theequation{\thesection.\arabic{equation}}}

\def\underline#1{\relax\ifmmode\@@underline#1\else
        $\@@underline{\hbox{#1}}$\relax\fi}

\def\titlepage{\@restonecolfalse\if@twocolumn\@restonecoltrue\onecolumn
     \else \newpage \fi \thispagestyle{empty}\c@page\z@
        \def\thefootnote{\fnsymbol{footnote}} }

\def\endtitlepage{\if@restonecol\twocolumn \else  \fi
        \def\thefootnote{\arabic{footnote}}
        \setcounter{footnote}{0}}  %\c@footnote\z@ }
\relax

\numberbysection
\hybrid

\def\beq{\begin{equation}}
\def\eeq{\end{equation}}
\def\p{\partial}
\def\G{\Gamma}
\def\g{\gamma}
\def\s{\sigma}
\def\L{{\cal L}}
\def\a{\alpha}
\def\b{\beta}

\def\D{{\cal D}}
\def\dim{{\rm dim}}
\def\res{{\rm res}}

\newtheorem{th}{Theorem}[section]

\begin{document}

\begin{titlepage}
\title{Elliptic solutions to difference non-linear equations and nested
Bethe ansatz equations}

\author{I.Krichever \thanks{Columbia University, 2990 Broadway,
New York, NY 10027, USA and
Landau Institute for Theoretical Physics,
Kosygina str. 2, 117940 Moscow, Russia; e-mail:
krichev@math.columbia.edu}}
\date{March 2, 1998}

\maketitle

\begin{abstract}

\noindent
We outline an approach to a theory of various generalizations of the
elliptic Calogero-Moser (CM) and Ruijsenaars-Shneider (RS) systems based on a
special inverse problem for linear operators with elliptic coefficients.
Hamiltonian theory of such systems is developed with the help of the
universal symplectic structure proposed by D.H. Phong and the author.
Canonically conjugated action-angle variables for spin generalizations of
the elliptic CM and RS systems are found.
\end{abstract}

\vfill

\end{titlepage}
\newpage

\section{Introduction}

The elliptic nested Bethe ansatz equations 
are a system of algebraic equations 
\beq 
\prod_{j\neq i}{\s(x_i^n-x_j^{n+1})\s(x_i^n-\eta-x_j^n)\s(x_i^n-x_j^{n-1}+\eta) 
\over \s(x_i^n-x_j^{n-1})\s(x_i^n+\eta-x_j^n)\s(x_i^n-x_j^{n+1}-\eta)}=-1 
\label{bet}
\eeq 
for $N$ unknown functions $x_i=x_i^n,\ i=1,\ldots, N$, of a discrete 
variable $n$ (\cite{wig}). (Here and below $\s(x)=\s(x|\omega_1, \omega_2)$,
$\zeta(x)=\zeta(x|\omega,\omega')$, and $\wp(x)=\wp(x|\omega,\omega')$ 
are the Weierstrass $\s$-, $\zeta$-, and $\wp$-functions corresponding to the 
elliptic curve with periods $2\omega,\ 2\omega'$.) This system is an example
of the whole family of integrable systems which have attracted renewed
interest for years. The most recent burst of interest is due to the unexpected 
connections of these systems to Seiberg-Witten solution of $N=2$ 
supersymmetric gauge theories \cite{sw}. It turns out that the low energy 
effective theory for $SU(N)$ model with matter in the adjoint representation
(identified first in \cite{dw} with $SU(N)$ Hitchin system) is isomorphic to 
the elliptic CM system. Using this connection quantum order parameters were
found in \cite{pd}.

The elliptic Calogero-Moser
(CM) system \cite{c}, \cite{m} is a 
system of $N$ identical particles on a line interacting with each other via 
the potential $V(x)=\wp(x)$. Its equations of motion have the form
\beq
\ddot x_i= 4\sum_{j\neq i}\wp'(x_i-x_j). \label{cm}
\eeq
The CM system is a completely integrable Hamiltonian system, i.e. it
has $N$ independent integrals $H_k$ in involution (\cite{op}, \cite{op1}). 
The second integral $H_2$ is the Hamiltonian of (\ref{cm}).

In \cite{amm} a remarkable connection of the CM system
with a theory of elliptic solutions to the KdV equation was revealed.
It was shown that the elliptic solutions of the KdV equations have the form
$
u(x,t)=2\sum_{i=1}\wp(x-x_i(t)) 
$
and the poles $x_i(t)$ of the solutions satisfy the constraint
$\sum_{j\neq i}\wp'(x_i-x_j)=0$,
which is the {\it locus} of the stationary points of the CM system. Moreover,
it turns out that the dependence of the poles with respect to $t$ coincides
with the Hamiltonian flow corresponding to the third integral $H_3$ of the
system. In \cite{kr3}, \cite{chood} it was found that this connection becomes 
an isomorphism in the case of the elliptic solutions to the 
Kadomtsev-Petviashvilii equation. Since then, the theory of the CM system and
its various generalizations is inseparable from the theory of the
elliptic solutions to the soliton equations.

In \cite{NRK} system (\ref{bet}) was revealed as the pole system 
corresponding to the elliptic solutions of completely discretized version 
of the KP equation on lattice. It was noticed that equations 
(\ref{bet}) have the form of the Bethe ansatz equations for the 
spin-${1\over 2}$\  Heisenberg chain with impurities. Its connection with the 
nested ansatz equations for the $A_k$-lattice models was established in 
\cite{wig}. 

As shown in \cite{zab} an intermediate discretization of the KP
equation which is the $2D$ Toda lattice equations leads to the
Ruijesenaars-Schneider system \cite{ruij}:
\beq 
\ddot x_i = \sum_{s\neq i} \dot
x_i \dot x_s (V(x_i-x_s)-V(x_s-x_i)), \ 
V(x)=\zeta(x)-\zeta(x+\eta),
\label{rs} 
\eeq
which is a relativistic version of (\ref{cm}).

The main goal of this paper is to present a general approach to the theory
of the CM type many body systems which is based on a special inverse problem 
for {\it linear} operators with the elliptic coefficients. 
This approach originated in 
\cite{kr1} and developed in \cite{wig}, \cite{bab}, \cite{zab}  clarifies the
connection of these systems with the soliton equations for which the
corresponding linear operator is the Lax operator. We formulate the inverse
problem in the next section and show that its solution is equivalent to
a finite-dimensional integrable system. We discuss an algebraic-geometric 
interpretation of the corresponding systems.

The advantage of our approach is that it generates the finite-dimensional
system simultaneously with its Lax representation. Until recently,
among its disadvantages was the missing connection to  the 
Hamiltonian theory. For example, in \cite{zab}  
spin generalization of the RS system was proposed and explicitly solved in 
terms of the Riemann theta-functions of auxiliary spectral curves. 
At the same time, all direct attempts to show that this system is Hamiltonian 
have failed, so far. 

One of the existing general approaches to the Hamiltonian theory of
the CM type systems is based on their geometric 
interpretation as reductions of geodesic flows on symmetric spaces \cite{op1}. 
Equivalently, these models can be obtained from free dynamics on a larger 
phase space possessing a rich symmetry by means of the Hamiltonian reduction 
\cite{KKS}. A generalization to infinite-dimensional phase spaces (cotangent 
bundles to current algebras and groups) was suggested in \cite{GN1}, \cite{GN2}. 
The infinite-dimensional gauge symmetry allows one to make the reduction to a
finite number of degrees of freedom. 

A further generalization of this approach consists in considering
dynamical systems on cotangent bundles to moduli spaces of stable
holomorphic vector bundles on Riemann surfaces. Such systems were
introduced by Hitchin in the paper \cite{Hi}, where their integrability
was proved. An attempt to identify the known many body integrable
systems in terms of the abstract formalism developed by Hitchin
was made in \cite{Nekras}. To do this, it is necessary to
consider vector bundles on algebraic curves with singular points.
It turns out that the class of integrable systems corresponding to
the Riemann sphere with marked points includes spin generalizations
of the CM model as well as integrable Gaudin magnets
\cite{Gaudin} (see also \cite{ER}). 

Unfortunately, a geometric interpretation of spin generalization
of the elliptic RS system has not been yet found. Recently, such realization 
and consequently the Hamiltonian theory of the rational degeneration of that
system were found in \cite{arut}.

In section 3 we develop Hamiltonian theory of the CM type systems
in the framework of the new approach to the Hamiltonian theory of soliton
equations proposed in \cite{kp1} and \cite{kp2}. It can be applied
evenly to any equation having the Lax representation. The symplectic structure
is constructed in terms of the Lax operator, only. We discuss
three basic examples: spin generalizations of the CM and the RS systems,
and the nested Bethe ansatz equations. We would like to emphasize that the 
universal form of the symplectic structure provides a universal and direct way 
to the action-angle type variables.

We would like to refer to \cite{pd} for the analysis of connections of the
elliptic CM system to Seiberg-Witten theory of $N=2$ supersymmetric gauge theories.

\section{Generating problem}

Let $\L$ be a linear differential or difference operator in 
two variables $x,t$ with coefficients which are scalar or matrix elliptic 
functions of the variable $x$ (i.e. meromorphic double-periodic functions
with the periods $2\omega_{\a}, \ \a=1,2$). We do not assume any special 
dependence of the coefficients with respect to the second variable. 
Then it is natural to introduce a notion of {\it double-Bloch} solutions 
of the equation 
\beq
\L \Psi=0. \label{gen}
\eeq
We call a {\it meromorphic} vector-function $f(x)$ that
satisfies the following monodromy properties:
\beq
f(x+2\omega_{\alpha})=B_{\alpha} f(x), \ \  \alpha=1,2,\label{g1}
\eeq
a {\it double-Bloch function}.  The complex numbers $B_{\alpha}$ are  called 
{\it Bloch multipliers}.  (In other words, $f$ is a meromorphic section of a 
vector bundle over the elliptic curve.)

In the most general form a problem that we are going to address is to
{\it classify} and to {\it construct} all the operators $\L$
such that equation (\ref{gen}) has {\it sufficiently enough} double-Bloch 
solutions. 

It turns out that existence of the double-Bloch solutions is so 
restrictive that only in exceptional cases such solutions do exist.  
A simple and general explanation of that is due to the Riemann-Roch
theorem. Let $D$ be a set of points $x_i,\ i=1,\ldots,m,$ on the elliptic 
curve $\G_0$ with multiplicities $d_i$ and let $V=V(D; B_1,B_2)$ be a linear 
space of the double-Bloch functions with the Bloch multipliers 
$B_\a$ that have poles at $x_i$ of order less or equal to $d_i$ and 
holomorphic outside $D$.
Then the dimension of $D$ is equal to:  
$$
{\rm dim} \ D={\rm deg} \ D=\sum_i d_i. 
$$
Now let $x_i$ depend on the variable $t$. Then for $f\in D(t)$ the function 
$\L f$ is a double-Bloch function with the same Bloch multipliers but in 
general with higher orders of poles because taking derivatives and  
multiplication by the elliptic coefficients increase orders. 
Therefore, the operator $\L$ defines a linear operator 
$$ 
\L|_D: V(D(t);B_1,B_2)\longmapsto V(D'(t);B_1,B_2), \ N'=\deg D'>N=\deg D, 
$$
and (\ref{gen}) is {\it always} equivalent to an {\it over-determined} 
linear system of $N'$ equations for $N$ unknown variables which are the 
coefficients $c_i=c_i(t)$ of an expansion of $\Psi\in V(t)$ with respect to 
a basis of functions $f_i(t)\in V(t)$. With some exaggeration one may say
that in the soliton theory the representation of a system  in the
form of the compatibility condition of an over-determined system of the
linear problems is considered as equivalent to integrability.

In all of the examples which we are going to discuss $N'=2N$ and
the over-determined system of equations has the form
\beq
LC=kC,\ \ \p_tC=MC,\label{lax}
\eeq
where $L$ and $M$ are $N\times N$ matrix functions depending on a point $z$
of the elliptic curve as on a parameter. A compatibility 
condition of (\ref{lax}) has the standard Lax form $\p_t L=[M,L]$, and is 
equivalent to a finite-dimensional integrable system. 

The basis in the space of the double-Bloch functions can be written in terms
of the fundamental function $\Phi(x,z)$ defined by the formula
\beq
\Phi(x,z)={\sigma(z-x)\over \sigma(z) \sigma(x)} 
e^{\zeta(z)x}. \label{phi}
\eeq
Note, that $\Phi(x,z)$ is a solution of the Lame equation:
\beq
\left({d^2\over dx^2}-2\wp(x)\right)\Phi(x,z)=\wp(z)\Phi(x,z). \label{lame}
\eeq
From the monodromy properties it follows that $\Phi$ considered
as a function of $z$ is double-periodic:
$$
\Phi(x,z+2\omega_{\alpha})=\Phi(x,z) ,
$$
though it is not elliptic in the classical sense due to an 
essential singularity at $z=0$ for $x\neq 0$.

As a function of $x$ the function $\Phi(x,z)$ is double-Bloch function, i.e.
$$
\Phi(x+2\omega_{\alpha}, z)=T_{\alpha}(z) \Phi (x, z), \ T_{\alpha}(z)=
\exp \left(2\omega_{\a}\zeta(z)-2\zeta (\omega _{\alpha})z\right).  
$$ 
In the fundamental domain of the lattice defined by
$2\omega_{\alpha}$ the function $\Phi(x,z)$ has a unique pole at the point 
$x=0$:  
\beq
\Phi(x,z)=x^{-1}+O(x). \label{j}
\eeq
The gauge transformation
$$
f(x)\longmapsto \tilde f(x)=f(x)e^{ax},
$$
where $a$ is an arbitrary constant does not change poles of any function and
transform a double Bloch-function into a double-Bloch function. If $B_{\alpha}$ 
are Bloch multipliers for $f$ than the Bloch multipliers for $\tilde f$ are 
equal to 
\beq
\tilde B_1=B_1e^{2a\omega_1},\ \ \tilde B_2=B_2 e^{2a\omega_2}. \label{gn2} 
\eeq
The two pairs of Bloch multipliers that are connected with each other
through the relation (\ref{gn2}) for some $a$ are called equivalent.
Note that for all equivalent pairs of Bloch multipliers the product
$
B_1^{\omega_2} B_2^{-\omega_1} 
$
is a constant depending on the equivalence class, only.

From (\ref{j}) it follows that a double-Bloch function $f(x)$ with simple 
poles $x_i$ in the fundamental domain and with Bloch multipliers 
$B_{\alpha}$ (such that at least one of them is not equal to $1$) may be 
represented in the form:  
\beq 
f(x)=\sum_{i=1}^N c_i\Phi(x-x_i,z) e^{k x},\label{g2} 
\eeq 
where $c_i$ is a residue of $f$ at $x_i$ and $z$, $k$ are parameters
related by 
\beq 
B_{\alpha}=T_{\alpha}(z) e^{2\omega_{\alpha}k }. \label{g3} 
\eeq 
(Any pair of Bloch multipliers may be represented in the form (\ref{g3}) 
with an appropriate choice of the parameters $z$ and $k$.)

To prove (\ref{g2}) it is enough to note that as a function of $x$
the difference of the left and right hand sides is holomorphic in the 
fundamental domain.  It is a double-Bloch function with the same Bloch 
multipliers as the function $f$. But a non-trivial double-Bloch function with 
at least one of the Bloch multipliers that is not equal to $1$, has at least 
one pole in the fundamental domain. 

\medskip

\noindent
Now we are in a position to present a few examples of the generating problem.

\noindent
{\bf Example 1. The elliptic CM system} (\cite{kr1}). 

Let us consider the equation
\beq 
\L\Psi=(\p_x-\p_x^2+u(x,t))\Psi=0 ,\label{cmo}
\eeq 
where $u(x,t)$ is an elliptic function. Then as shown in \cite{kr1} equation
(\ref{cmo}) has $N$ linear independent double-Bloch solutions with equivalent 
Bloch multipliers and $N$ simple poles at points $x_i(t)$ if
and only if $u(x,t)$ has the form 
\beq
u(x,t)=2\sum_{i=1}^N\wp(x-x_i(t))     \label{u}
\eeq
and $x_i(t)$ satisfy the equations of motion of the elliptic CM system 
(\ref{cm}).

The assumption that there exist $N$ linear independent double-Bloch
solutions with equivalent Bloch multipliers implies that they can be
written in the form
\beq
\Psi=\sum_{i=1}^N c_i(t,k,z)\Phi(x-x_i(t),z)e^{kx+k^2t}, \label{psi}
\eeq
with the same $z$ but different values of the parameter $k$.

Let us substitute (\ref{psi}) into (\ref{cmo}). Then 
(\ref{cmo}) is satisfied if and if we get 
a function holomorphic in the fundamental domain. First of all, we
conclude that $u$ has poles at $x_i$, only. 
The vanishing of the triple poles $(x-x_i)^{-3}$ implies
that $u(x,t)$ has the form (\ref{u}). The vanishing of the double
poles $(x-x_i)^{-2}$ gives the equalities that can be written as a matrix 
equation for the vector $C=(c_i)$:  
\beq (L(t,z)-kI)C=0\,,
\label{l} 
\eeq
where $I$ is the unit matrix and the Lax matrix $L(t,z)$ is defined as 
follows
\footnote{in order to simplify the consequent formulae in Section 3
and present all the examples in the same framework we choose here and in the 
next example a normalization of $L$ which differs from that used in
\cite{kr1}, \cite{bab} by the factor $-1/2$.}
:  
\beq 
L_{ij}(t,z)=-{1\over 2}\delta_{ij}\dot x_i-(1-\delta_{ij})\Phi(x_i-x_j,z).  
\label{L} 
\eeq 
Finally, the vanishing of the simple poles gives the equations 
\beq (\partial_t-M(t,z))C=0\,,  \label{m} \eeq where \beq
M_{ij}=\left(\wp(z)-2\sum_{j\neq 
i}\wp(x_i-x_j)\right)\delta_{ij}-2(1-\delta_{ij})\Phi' (x_i-x_j ,z). \label{M} 
\eeq 
The existence of $N$ linear independent 
solutions for (\ref{cmo}) with equivalent Bloch multipliers implies that 
(\ref{l}) and (\ref{m}) have $N$ independent solutions corresponding to 
different values of $k$. Hence, as a compatibility condition we get the Lax 
equation $\dot L=[M,L]$ which is equivalent to (\ref{cm}). Note that the last 
system does not depend on $z$. Therefore, if (\ref{l}) and (\ref{m}) are 
compatible for some $z$ then they are compatible for all $z$. As a result we 
conclude that if (\ref{cmo}) has $N$ linear independent double-Bloch solutions 
with equivalent Bloch multipliers then it has infinitely many of them. All the 
double-Bloch solutions are parameterized by points of an algebraic curve $\G$ 
defined by the characteristic equation 
\beq 
R(k,z)\equiv \det (kI-L(z))=k^N+\sum_{i=1}^Nr_i(z)k^{N-i}=0. \label{r} \eeq 
Equation (\ref{r}) can be seen as a dispersion relation between two Bloch 
multipliers and defines $\G$ as $N$-sheet cover of $\G_0$. 

From (\ref{phi}) and (\ref{L}) it follows that 
\beq
L=G\widetilde L G^{-1}, \ \ G_{ij}=e^{\zeta(z)x_i}\delta_{ij}. \label{cal}
\eeq
At $z=0$ we have 
the form 
\beq 
\widetilde L=z^{-1}(F-I)+O(1), \ \ F_{ij}=1. \label{10}
\eeq 
The matrix $F$ has zero eigenvalue with multiplicity $N-1$ and a
simple eigenvalue $N$. Therefore, in a neighborhood of $z=0$ 
the characteristic polynomial (\ref{r}) 
has the form 
\beq
R(k,z)=\prod_{i=1}^N(k+\nu_iz^{-1}+h_i+O(z)),\ \ \nu_1=1-N, \ \nu_i=1, \ i>1. 
\label{r1}
\eeq
We call the sheet of the covering $\G$ at $z=0$ corresponding to the branch 
$k=z^{-1}(N-1)+O(1)$ by upper sheet and mark the point $P_1$ on this 
sheet among the preimages of $z=0$. From (\ref{r1}) it follows that in general 
position when the curve $\G$ is smooth, its genus equals $N$.

The coefficient $r_i(z)$ in (\ref{r}) is an elliptic function with a pole at 
$z=0$ of order $i$. The coefficients $I_{i,s}$ of the expansion 
\beq 
r_i(z)=I_{i,0}+\sum_{s=0}^{i-2}I_{i,s}\p_z^s\wp(z) 
\eeq 
are integrals of motion. From (\ref{r1}) it is easy to show that there are 
only $N$ independent among them. Recently, a remarkable explicit
representation for the characteristic equation (\ref{r}) was found 
(\cite{pd}:  
\beq 
R(k,z)=f(k-\zeta(z),z), \ \ f(k,z)={1\over \s(z)}\ \s
\left(z+{\p\over \p k}\right)H(k),
\label{ph}
\eeq
where H(k) is a polynomial.
Note that (\ref{ph}) may be written as:
$$
f(k,z)={1\over \s(z)}\ \sum_{n=1}^N {1\over n!}\p_z^n \s(z)
\left({\p\over \p k}\right)^n H(k).
$$
The coefficients of the polynomial $H(k)$ are free parameters of the spectral 
curve of the CM system.

When the Lax representation and the corresponding algebraic curve $\G$
are constructed, the next step is to consider analytical properties
of the eigenvectors of the Lax operator $L$ on $\G$. 
   
Let $L$ be a matrix of the form (\ref{L}). Then the components of the 
eigenvector $C(P)=(c_i(P)), \ P=(k,z)\in \G,$ normalized by the condition 
\beq
\sum_{i=1}^Nc_i(P)\Phi(-x_i,z)=1, \label{nor} 
\eeq
are meromorphic functions on $\G$ outside the preimages $P_{i}$ of $z=0$.  
They have $N$ poles $\g_1,\ldots,\g_N$ (in a general position $\g_s$ are 
distinct). At the points $P_j$ the coordinates $c_i(P)$ have the expansion:  
\beq
c_i=z(c_{i}^1+O(z))e^{\zeta(z)x_i};\ 
c_i=(c_{i}^j+O(z))e^{\zeta(z)x_i}, \ i>1. 
  \eeq
If we denote the diagonal elements of $L$ by $-p_i/2$, then the above
constructed correspondence
\beq 
p_i, \ x_i\longmapsto \{\G,\D=\{\g_s\}\} \label{cor}
\eeq 
is an isomorphism (on the open set).

Now let $x_i(t)$ be a solution of (\ref{cm}) then the divisor $\D$ 
corresponding
to $p_i=\dot x_i(t), \ x_i(t)$ depends on $t,\ \D=\D(t)$. It turns out that 
under the
Abel transform this dependence becomes linear on the Jacobian $J(\G)$ of
the spectral curve. The final result is as follows. 
\begin{th}
The coordinates of the particles $x_i(t)$ are roots of the equation 
\beq
\theta(Ux+Vt+Z_0)=0, \label{for} 
\eeq
where $\theta(\xi)=\theta(\xi|B)$ is the Riemann theta-function 
corresponding to matrix of $b$-periods of holomorphic differentials on $\G$; 
the vectors $U$ and $V$ are the vectors of $b$-periods of normalized 
meromorphic differentials on $\G$, with poles of order 2 and 3 at the point 
$P_1$; the vector $Z_0$ is the Abel transform of the divisor $\D_0=\D(0)$.
\end{th}
This result can be reformulated in the following more geometric way.
Let $J(\G)$ be the Jacobain ($N$-dimensional complex torus) of a smooth genus 
$N$ algebraic curve $\G$. Abel transform defines 
imbeding of $\G$ into $J(\G)$. A point $P\in \G$ defines a vector $U$ in 
$J(\G)$  that is the tangent vector to the image of $\G$ at the point. 
Let us consider a class of curves having the following property: there exists a point on the 
curve such that the complex linear subspace generated by the corresponding 
vector $U$ {\it is compact}, i.e. it is an elliptic curve $\G_0$.  This means 
that there exist two complex numbers $2\omega_{\alpha}, \ {\rm Im} \ 
\omega_2/\omega_1>0$, such that $2\omega_{\alpha}U$ belongs to the lattice of 
periods of the holomorphic differentials on $\G$.  From pure 
algebraic-geometrical point of view the problem of the description of such  
curves is transcendental. It turns out that this problem has an explicit 
solution, and algebraic equations that define such curves are the 
characteristic equation for the Lax operator corresponding to the CM system.  
Moreover, it turns out that in general position  $\G_0$ intersects 
theta-divisor at $N$ points $x_i$ and if we move $\G_0$ in the direction that 
is defined by the vector $V$ of the second jet of $\G$ at $P$ then the 
intersections of $\G_0$ with the theta-divisor move according to the CM 
dynamics.   

\medskip

\noindent{\bf  Example 2. Spin generalization of the elliptic CM system} 
(\cite{bab}).

Let $\L$ be an operator of the same form (\ref{cmo}) as in the previous case,
but now $u=u_{\a}^{\b}(x,t)$ is an elliptic $(l\times l)$ {\it matrix} 
function of the variable $x$. We slightly reformulate the results of \cite{bab} 
to a form that would be used later.

Equation (\ref{cmo}) has $N\geq l$ linear independent double-Bloch solutions 
with $N$ simple poles at points $x_i(t)$ and such that $(l\times N)$ matrix 
formed by its residues at the poles has rank $l$ if and only if:\\ 
(i) the potential $u$ has the form 
\beq 
u=\sum_{i=1}^N a_i(t)b_i^+(t) 
\wp(x-x_i(t)),\label{u1} 
\eeq 
where $a_i=(a_{i,\a})$ are $l$-dimensional vectors and $b_i^+=(b_i^{\a})$ are 
$l$-dimensional co-vectors;\\ 
(ii) $x_i(t)$ satisfy the equations 
\beq 
\ddot x_i=\sum_{j\neq i}(b_i^+a_j)(b_j^+a_i)\wp'(x_i-x_j); \label{scm} 
\eeq 
(iii) the vectors $a_i$ and co-vectors $b_i^+$ satisfy the constraints 
\beq
b_i^+a_i=\sum_{\a=1}^l b_{i}^{\a}(t)a_{i\a}(t) =2 \label{cons} \eeq
and  the equations 
\beq
\dot{a}_i= -\sum_{j\neq i}a_j(b_j^+a_i)\wp(x_i-x_j)-\lambda_i a_i,\ \ 
\dot{b}_i=\sum_{j\neq i}b_j(b_i^+a_j)\wp(x_i-x_j)+ \lambda_ib_i^+
\label{ab} 
\eeq
where $\lambda_i=\lambda_i(t)$ are scalar functions. 

In order to get these results we represent a double-Bloch vector
function $\Psi$ in the form:
\beq
\Psi=\sum_{i=1}^N s_i(t,k,z)\Phi(x-x_i(t),z)e^{kx+k^2t}, \label{Psi}
\eeq
where $s_i$ are $l$-dimensional vectors and substitute it into (\ref{cmo}). 
The vanishing of the triple pole $(x-x_i)^{-3}$ implies that $u$ has the form
(\ref{u1}), the vectors $s_i$ are proportional to $a_i$, i.e.
$s_i=c_ia_i$, where $c_i$ are scalars, and that the constraints (\ref{cons})
are fulfilled.

The vanishing of the coefficients in front of $(x-x_i)^{-2}$ implies that 
the vector $C$ with the coordinates $c_i$ satisfies (\ref{l}), where the 
Lax  matrix has the form
\beq
L_{ij}=-{1\over 2}\delta_{ij}\dot x_i -
{1\over 2}(1-\delta_{ij}) (b_i^+a_j)\Phi(x_i-x_j,z). \label{L1} 
\eeq
The vanishing of the coefficients in front of $(x-x_i)^{-1}$ implies (\ref{ab}) 
for the vectors $a_i$ and the equation (\ref{m}), where the matrix $M$ has the 
form 
\beq 
M_{ij}=(\lambda_i+\wp(z))\delta_{ij}-(1-\delta_{ij})(b_i^+a_j)\Phi'(x_i-x_j). 
\label{M1} 
\eeq 
The Lax equation for these matrices implies (\ref{scm}) and 
equations (\ref{ab}) for $b_i^+$ (here we use the assumption that $a_i$ span the 
whole $l$-dimensional space).

System (\ref{ab}) after the gauge transformation
\beq 
a_i\to a_iq_i,
\ b_i^+\to b_iq_i^{-1},\ \ q_i=\exp\left(\int^t \lambda_i(t)dt\right), 
\label{1.10} 
\eeq 
which does not effect (\ref{scm}) and (\ref{cons}), becomes
\beq 
\dot{a}_i= -\sum_{j\neq i}a_j(b_j^+a_i)\wp(x_i-x_j),\ \
\dot{b}_i=\sum_{j\neq i}b_j(b_i^+a_j)\wp(x_i-x_j). \label{AB}
\eeq Equations (\ref{cons}), (\ref{AB}) are invariant under the transformations
\beq 
a_i\to \lambda_i^{-1} a_i,\  \ b_i^+\to \lambda_i b_i^+,\ \
a_i\to W^{-1}a_i, \ \ b_i^+\to b_i^+W, \label{20} 
\eeq
where $\lambda_i$ are constants and $W$ is a constant $(l\times l)$ matrix. A 
factorization with respect to these transformations leaves us with a reduced 
phase space ${\cal M}$ of the dimension ${\rm dim} \ {\cal M}=2Nl-l(l-1)$.
Let us introduce a canonical system of coordinates on ${\cal M}$.  
First of all, 
for any set of $a_i$, $b_i^+$ we define the matrix 
\beq
S_{\a}^{\b}=\sum_{i=1}^N a_{i\a}b_i^{\b} \label{S} 
\eeq
and diagonalize it with the help of the matrix $W_{0\a}^j$ which leaves the 
co-vector $(1,\ldots,1)$ invariant, i.e.  
\beq
S_{\a}^{\b}W_{0\b}^j=2\kappa_jW_{0\a}^j,\ \ \sum_{\a}W_{0\a}^j=1, \ 
j=1,\ldots,l.  \label{W} 
\eeq
Then we define 
\beq
A_i=W_0^{-1}a_i,\ B_i^+=b_i^+W_0. \label{can} 
\eeq
The vectors $A_i$ and co-vectors $B_i$ satisfy the conditions
\beq
\sum_{i=1}^N A_{i\a}B_i^{\b}=2\kappa_{\a}\delta_{\a}^{\b}, \label{n} 
\eeq
which destroy the second half of the gauge transformations (\ref{20}).

At $z=0$ we have
\beq
L=G\widetilde LG^{-1}, \ \widetilde L=z^{-1}(F-I)+O(1), \ \ 
F_{ij}={1\over 2}(b_i^+a_j)={1\over 2}(B_i^+A_j), \label{F}
\eeq
where $G$ is the same as in (\ref{cal}).
The matrix $F$ has rank $l$. Its null subspace is a subspace of vectors
such that
\beq
\sum _{i=1}^N A_{i,\a}c_i=0. \label{null}
\eeq
Relations (\ref{n}) imply that the eigenvector of $F$ corresponding to 
non-zero eigenvalue $2\kappa_j$ is identified with $B_i^{j}$.

From (\ref{F}) it follows that
\beq
R(k,z)=\prod_{i=1}^N(k+\nu_iz^{-1}+h_i+O(z)), \ \nu_i=1-\kappa_i,\ i\leq l,
\ \nu_i=1,\ i>l. \label{21} 
\eeq 
As shown in \cite{bab} expansion (\ref{21}) 
implies that the spectral curve defined by (\ref{r}) has (in general position) 
genus $g=Nl-l(l+1)/2+1$.  At the same time (\ref{21}) implies that a 
number of independent integrals given by (\ref{r}) is equal to 
${1\over 2} \dim \ {\cal M}$.

The angle-type variables of our reduced system are the divisor
of poles of the solution of (\ref{l}) with the following
normalization:
\beq
\sum_{\a=1}^l\sum_{i=1}^N A_{i\a}c_i\Phi(-x_i,z)=1. \label{nor1}
\eeq
At the points $P_j$ the components of $C$ has the form
\beq
c_i=z(c_{i}^j+O(z))e^{\zeta(z)x_i},\ i\leq l;\ 
c_i=(c_{i}^{j}+O(z))e^{\zeta(z)x_i},\ j>l, \label{22}
\eeq
where
\beq
\sum _{i=1}^N A_{i,\a}c_i^j=\delta _{\a}^j,\ j\leq l,\ \ 
\sum _{i=1}^N A_{i,\a}c_i^j=0, \ j>l. \label{nll}
\eeq
The last formulae identify the normalization (\ref{nor1}) with a canonical
normalization used in the soliton theory. The coordinates
$$
\Psi_{\a}=\sum_{i=1}^N A_{i\a}c_i \Phi (x-x_i,z)^{kx}
$$
of the corresponding
Baker-Akhiezer function $\Psi$ are  meromorphic outside the punctures 
$P_j,\ j\leq l$. At $P_j$ they have  the form:
$$
\Psi_{\a}=\left(\delta _{\a}^j +O(z)\right) e^{z^{-1}\kappa_j x}
$$
(see details in \cite{bab}).

Outside the punctures $P_j$ the canonically normalized vector $C$
has $g+l-1$ poles $\g_1,\ldots,\g_{g-l+1}$. This number is equal
to ${1\over 2}\ \dim \ {\cal M}$. Note that these poles are independent
on the transformations (\ref{20}). Therefore, we have constructed
an algebraic-geometric correspondence
\beq
{\cal M}=\{x_i,\ p_i, \ A_i,\ B_i\}\longmapsto \{\G,\D=(\g_s)\}\ , \label{30}
\eeq
which is an isomorphism on the open set. 
The reconstruction formulae for solutions of the elliptic CM system
(\ref{scm}), (\ref{cons}), (\ref{AB})  can be found in \cite{bab}.
It turns out that the coordinates of $x_i$ are defined by the
same equation (\ref{for}). The only difference is a set of the corresponding
algebraic curves that are defined by the characteristic equation for the Lax
matrix $L$ of the form (\ref{L1}). 
In pure algebraic-geometric form they can be described
in a way similar to that in the previous example. Namely, they are curves
such that there exists a set of $l$ points on it with the following property:
a linear space spanned by the tangent vectors to the curve at these points in
the Jacobian, contains a vector $U$ which spans in $J(\G)$ an elliptic curve
$\G_0$.

\medskip

\noindent{\bf  Example 3. Spin generalization of the elliptic RS system}
(\cite{zab})

Let us consider now the differential-difference equation
\beq
\L\Psi=\p_t\Psi(x,t)-\Psi(x+\eta,t)-v(x,t)\Psi(x,t)=0, \label{41}
\eeq
where $\eta$ is a complex number and $v(x,t)$ is an elliptic
$(l\times l)$ matrix function. It has $N\geq l$ linear independent
double-Bloch solutions with $N$ simple poles at points $x_i(t)$ and such that
$(l\times N)$ matrix formed by its residues at the poles has rank $l$
if and only if:\\ 
(i) the potential $u$ has the form
\beq
v=\sum_{i=1}^N a_i(t)b_i^+(t) V(x-x_i(t)), \ 
V=\zeta(x-x_i+\eta)-\zeta(x-x_i),\label{42} 
\eeq 
where $a_i=(a_{i,\a})$ are $l$-dimensional vectors and $b_i^+=(b_i^{\a})$ 
are $l$-dimensional co-vectors;\\
(ii) $x_i(t)$ satisfy the equations 
\beq
\ddot x_i=\sum_{j\neq i}(b_i^+a_j)(b_j^+a_i)\left(V(x_i-x_j)-V(x_j-x_i)\right); 
\label{43}
\eeq
(iii) the vectors $a_i=(a^i_{\a})$ and co-vectors $b_i=(b_i^{\a})$ satisfy the 
constraints 
\beq 
b_i^+a_i=\sum_{\a=1}^l b_{i}^{\a}(t)a_{i\a}(t) =\dot x_i \label{44}
\eeq
and  the system of equations 
\beq
\dot{a}_i= \sum_{j\neq i}a_j(b_j^+a_i)V(x_i-x_j)-\lambda_i a_i,\ \ 
\dot{b}_i=-\sum_{j\neq i}b_j(b_i^+a_j)V(x_j-x_i)+ \lambda_ib_i^+,
\label{45} 
\eeq
where $\lambda_i=\lambda_i(t)$ are scalar functions. 

The gauge transformation (\ref{20}) allows us to eliminate $\lambda_i$ in
(\ref{45}). The corresponding system was introduced in \cite{zab} and
is spin generalization of the elliptic RS model, 
which coincides with (\ref{43}), (\ref{44}) for $l=1$.

Construction of integrals of system (\ref{43})-(\ref{45}) 
and the angle type coordinates is parallel to the previous case but 
requires some technical modification
because $\L$ is a difference operator in $x$. First of all, we choose
a different basis in a space of double-Bloch functions. The proper
choice is defined by the formula
\beq
\Phi(x,z)={\sigma(z+x+\eta)\over \sigma(z+\eta) \sigma(x)} \left[
{\sigma(z-\eta)\over \sigma(z+\eta)}\right]^{x/2\eta}. \label{1.5}
\eeq
It satisfies the difference analog of the Lame equation (\ref{lame}):
$$
\Phi(x+\eta,z)+c(x)\Phi(x-\eta,z)=E(z)\Phi(x,z) , 
$$
where
$$
c(x)={\sigma(x-\eta)\sigma(x+2\eta)\over \sigma(x+\eta)\sigma(x)} ,\ \ \ 
E(z)={\sigma(2\eta)\over \sigma(\eta)} {\sigma(z)\over
(\sigma(z-\eta)\sigma(z+\eta))^{1/2}}. 
$$
The Riemann surface $\hat {\G}_0$ of the function $E(z)$
is a two-fold covering of the initial elliptic curve
$\G_0$ with periods $2\omega_{\alpha}, \
\alpha=1,2$. Its genus is equal to $2$.

As a function of $x$ the function $\Phi(x,z)$ is double-Bloch function.
In the fundamental domain of the lattice defined by
$2\omega_{\alpha}$, the function $\Phi(x,z)$ has a unique
pole at the point $x=0$:
\beq
\Phi(x,z)=x^{-1}+A+ O(x), \ \ A=\zeta (z+\eta )+\frac{1}{2\eta }
\ln \frac{\sigma (z-\eta )}{\sigma(z+\eta )}. \label{1.12}
\eeq
Therefore, we may represent a double-Bloch solution $\Psi$ of (\ref{41})
in the form:
\beq
\Psi=\sum_{i=1}^N s_i(t,k,z)\Phi(x-x_i(t),z) k^{x/\eta},\label{1.7}
\eeq
substitute this ansatz into the equation and proceed as before. We get 
(\ref{43})-(\ref{45}). The corresponding Lax operators have the form
\beq
L_{ij}(t,z)=(b_{i}^{+}a_j)\Phi(x_i-x_j-\eta,z),  \label{LaxL}
\eeq
\beq
M_{ij}(t,z)=(\lambda_i -(\zeta
(\eta)-A)\dot{x}_i )\delta_{ij}+
(1-\delta_{ij})(b_{i}^{+}a_j) \Phi (x_i -x_j ,z). \label{LaxM}
\eeq
Explicit formulae in terms of the Riemann theta functions are the same
as for spin generalization of the CM system. The only difference is due
to the different family of the spectral curves. 
In the case $l=1$ they may be defined as a class of curves
having the following property: there exists
a pair of points on the curve such that the complex linear subspace
spanned by the corresponding vector $U$ is an elliptic curve $\G_0$. 
If we move $\G_0$ in the direction that is defined by the vector 
$V^+$ ($V^-$) tangent to $\G\in J(\G)$ at the 
point $P^+$ ($P^-$), then the intersections $x_i$ of $\G_0$ with the 
theta-divisor move according to the RS dynamics. The spectral curves
for $l>1$ are characterized by the existence of two sets of points 
$P_i^{\pm}, \ i=1,\ldots,l$ such that in the linear subspace spanned by the 
vectors corresponding to each pair there exists a vector $U$ with the same 
property as above.

\newpage

\noindent
{\bf Example 4. The nested Bethe ansatz equations} (\cite{wig})
 
Let us consider two-dimensional difference equation
\beq
\Psi(x,m+1)=\Psi(x+\eta)+v(x,m)\Psi(x,m) \label{51}
\eeq
with an elliptic in $x$ coefficient of the form
\beq
v(x,m)=\prod_{i=1}^N{\s(x-x_i^m)\s(x-x_i^{m+1}+\eta)\over
\s(x-x_i^{m+1})\s(x-x_i^{m}+\eta)}\ . \label{52}
\eeq
It has $N$ linear independent double-Bloch solutions with equivalent 
Bloch multipliers if and only if the functions $x_i^m$ of the discrete
variable $m$ satisfy the nested Bethe ansatz equations. 
The Lax representation for this system has the form
\beq
L(m+1)M(m)=M(m)L(m+1),
\eeq
where 
\beq
L_{ij}(m)=\lambda _{i}(m) \Phi(x_i^{m} - x_j^m -\eta ,z),\ 
M_{ij}(m)=\mu _{i}(m)\Phi(x_i^{m+1}- x_j^m , z) \label{LM}
\eeq
and 
\beq
\lambda _{i}(m)=\frac{\prod_{s=1}^{M}
\sigma(x_i^m- x_s^{m}-\eta )\sigma(x_i^m - x_s^{m+1})}
{\prod _{s=1, \ne i}^{M}\sigma(x_i^m - x_s^{m})
\prod _{s=1}^{N}\sigma(x_i^m - x_s^{m+1}-\eta )}\,,
\label{L2}
\eeq
\beq
\mu _{i}(m)=\frac{
\prod_{s=1}^{M}
\sigma(x_i^{m+1}- x_s^{m+1}+\eta )\sigma(x_i^{m+1} - x_s^{m})}
{\prod _{s=1, \ne i}^{M}\sigma(x_i^{m+1} - x_s^{m+1})
\prod _{s=1}^{N}\sigma(x_i^{m+1} - x_s^{m}+\eta )}\,.
\label{M2}
\eeq
A class of the spectral curves is the same as for the RS system. 
The solution $x_i^m$ of (\ref{bet}) corresponding to the
spectral curve and the divisor on it is defined by the equation
\beq
\theta(Ux+Vm+Z)=0. 
\eeq
Here $V$ is the vector from the puncture $P_+$
to the third point $Q$ on $\G$. When this point tends to $P_+$ the vector
$V$ becomes  the tangent vector to the curve and we come to the RS system
as a continuous limit of (\ref{bet}).

\section{Hamiltonian theory of the CM type systems.}

As we have seen various generating problems
lead to various integrable finite-dimensional systems which
can be explicitly solved via the spectral transform of a phase
space $\cal M$ to algebraic-geometric data. 
On this way we do not use Hamiltonian description of the system.
Moreover, a'priori it's not clear, why all the systems
which can be constructed with the help of the generating scheme
are Hamiltonian. In this section we clarify this problem using
the approach to the Hamiltonian theory of soliton equations
proposed in \cite{kp1} and developed in \cite{kp2}. 
 
First of all, let us outline a framework that was presented in the
previous section. The direct spectral transform identifies a space
of solutions with a bundle over  a space of the corresponding
spectral curves. The fiber over a curve $\G$ is a symmetric power
$S^{g+l-1}\G$ (i.e. an unordered set of $(g+l-1)$ points $\g_s\in \G$). 
Dimension of the space of the spectral curves equals to 
$g+l-1={1\over 2}\ {\cal M}$. The spectral curves  are realized as 
$N$-sheet covering of the elliptic curve $\G_0$.

Let us consider the case of spin generalization of the CM system. Entries 
of $L(z)$ are explicitly defined as functions on $\cal M$. 
Therefore, $L(z)$ can be seen as an operator-valued function 
and its external differential $\delta L$ as an operator-valued one-form
on $\cal M$. Canonically normalized eigenfunction $C(z,k)$ of $L(z)$ is 
the vector-valued function on $\cal M$. Hence, its differential is
a vector-valued one-form. Let us define a two-form on ${\cal M}$ 
by the formula
\beq
\omega=\sum_{i=1}^N\res_{P_i} <C^*(z,k) (\delta L(z)
+\delta k)\wedge \delta C(z)>dz,  \label{54}
\eeq
where $C^*(z,k)$ is the eigen-covector (row vector) of $L(z)$, i.e.
the solution of the equation $C^*L=kC^*$, normalized by the condition
$<C^*(z,k)C(z,k)>=1$.
The form $\omega$ can be rewritten as
\beq
\omega=\res_{z=0} \ {\rm Tr}\left(\hat C^{-1}(z) \delta L(z)\wedge \delta 
\hat C(z)-\hat C^{-1}(z) \delta \hat C(z)
\wedge \delta \hat k\right) dz,
\label{55}
\eeq
where $\hat C(z)$ is a matrix with columns $C(z,k_j)$;
$k_j=k_j(z)$ are different eigenvalues of $L(z)$ and $\hat k(z)$
is the diagonal matrix $k_j(z)\delta_{ij}$.

\noindent{\it Remark 1.} The right hand side of formula (\ref{54}) is not 
gauge invariant. In \cite{kp1} a gauge ($C_1=gC, \ L_1=gLg^{-1}$) was chosen 
in such a way that $\omega$ is equal to the sum of residues of the 
differential $<C_1^*\delta L_1 \wedge \delta C_1>dz$.

Note that $C^*$ are rows of the matrix $\hat C^{-1}$. That implies that $C^*$ as a 
function on the spectral curve is: meromorphic outside the punctures; has poles 
at the branching points of the spectral curve, and zeros at the poles $\g_s$ of 
$C$. These analytical properties are used in the proof of the following theorem.  
\begin{th} The two-form $\omega$ equals 
\beq 
\omega=2\sum_{s=1}^{g+l-1}\delta z(\g_s)\wedge \delta k(\g_s). \label{61} 
\eeq 
\end{th}  
The meaning of the right hand side of this formula is as follows. The spectral 
curve by definition arises with the meromorphic function $k(Q)$ and multi-valued 
holomorphic function $z(Q)$. Their evaluations $k(\g_s),\ z(\g_s)$ at the points 
$\g_s$ define functions on the space $\cal M$, and the wedge product of their 
external differentials is a two-form on $\cal M$.  Formula (\ref{61}) identifies 
$k_s=k(\g_s),\ z_s=z(\g_s)$ as Darboux coordinates for $\omega$.

\medskip
\noindent{\it Remark 2}. The right hand side of (\ref{61}) can be identified
with a particular case of universal algebraic-geometric symplectic forms
proposed in \cite{kp1}. They are defined on the generalized Jacobian bundles
over a proper subspaces of the moduli spaces of Riemann surfaces with
punctures. In the case of families of hyperelliptic curves that form was 
pioneered by Novikov and Veselov \cite{nv}.

\medskip
\noindent{\it Remark 3}.
Equations (\ref{scm}), (\ref{cons}), (\ref{AB}) are linearized by generalized
Abel transform
\beq
A: S^{g+l-1}\G \longmapsto J(\G)\times C^{l-1} . \label{A}
\eeq
This transform is defined with the help of a basis of the normalized 
holomorphic differentials $d\omega_i, \ i\leq g,$ and with the help of
normalized meromorphic differentials $d\Omega_j,\ 
j\leq (l-1)$ of the third kind with residues $1$ and $-1$ at
the points $P_j$ and $P_l$, respectively. Let $Q$ be a point of $\G$. 
Then we define $(g+l-1)$-dimensional vector $A_k(Q)$ with the
coordinates
$$
A_i(Q)=\int^Q d\omega_i,\ \ A_{g+j}(Q)=\int^Q d\Omega_j .
$$
The isomorphism (\ref{A}) is given by 
$$
\phi_k=\sum_{s=1}^{g+l-1} A_k(\g_s) .
$$
The action variables $I_k$ canonically conjugated to $\phi$, i.e. such that
$$
\omega=\sum_{k=1}^{g+l-1} \delta \phi_k \wedge \delta I_k,\ 
$$
are equal to
$$
I_i=\oint_{a_i^0} \ kdz ,\ I_{g+j}= \res_{P_j} \ kdz= -\nu_j,
$$
where $a_i^0$ are $a$-cycles of the basis of cycles on $\G$ with
the canonical matrix of intersections.

\medskip

Let us outline the proof of Theorem 3.1. The differential
$\Omega=<C^*\delta L\wedge \delta C>dz$ is a meromorphic differential
on the spectral curve (the essential singularities of the factors cancel each
other at the punctures). Therefore, the sum of its residues at the punctures 
is equal to the sum of other residues with negative sign.
There are poles of two types. First of all, $\Omega$ has poles at the poles
$\g_s$ of $C$. Note that $\delta C$ has pole of the second order
at $\g_s$. Taking into account that $C^*$ has zero at $\g_s$ we obtain
\beq
\res_{\g_s}\Omega=<C^*\delta LC>\wedge \delta z(\g_s)=\delta 
k(\g_s)\wedge\delta z(\g_s). \label{65}
\eeq
The last equality follows from the standard formula for a variation of
the eigenvalue of an operator. The second term in (\ref{54}) has
the same residue at $\g_s$. 

The second set of poles of $\Omega$ is a set of the branching points 
$q_i$ of the cover. The pole of $C^*$  at $q_i$ cancels with the zero 
of the differential $dz, \ dz(q_i)=0$, considered as differential on $\G$.  
The function $C$ is holomorphic at $q_i$. If we take an expansion of $C$ in 
the local coordinate $(z-z(q_i))^{1/2}$ (in general position when the 
branching point is simple) and consider its variation we get that
\beq 
\delta C=-{dC\over dz}\delta z(q_i)+O(1).\label{66}
\eeq 
Therefore, $\delta C$ has simple pole at $q_i$. In the similar way
we obtain 
\beq
\delta k=-{dk\over dz} \delta z(q_i). \label{67} 
\eeq
Equalities (\ref{66}) and (\ref{67}) imply that 
\beq
\res_{q_i}\Omega=\res_{q_i}\left[ <C^*\delta L dC>\wedge {\delta k dz\over 
dk}\right]\ . \label{68} 
\eeq
Due to skew-symmetry of the wedge product we we may replace $\delta L$ in 
(\ref{68}) by $(\delta L-\delta k)$. Then using identities 
$C^*(\delta L-\delta k)= \delta C^* (k-L)$  and 
$(k-L)dC=(dL-dk)C$ we obtain
\beq 
\res_{q_i}\Omega=-\res_{q_i}<\delta C^*C>\wedge \delta k dz=
\res_{q_i}<C^*\delta C>\wedge \delta k dz. \label{000}\eeq
(Note, that the term with $dL$ does not contributes to the residue, because
$dL(q_i)=0$.) The right hand side of (\ref{000}) cancels with a residue of 
the second term in the sum (\ref{54}).  The theorem is proved.

Now let us express $\omega$ in terms of the coordinates (\ref{can})
on $\cal M$. Using the gauge transformation 
\beq
L=G\ \widetilde L\ G^{-1},\ \ \hat C=G\ \widetilde C  \label{69}
\eeq
where $G$ is given by (\ref{cal}), 
we obtain
\beq
\omega=\sum_j\res_{z=0}\ {\rm Tr}\left[\delta \widetilde L\wedge \delta h+
\widetilde C^{-1}\left( \delta \widetilde L\wedge \delta \widetilde C +
[\delta h, \widetilde L] \wedge \delta \widetilde C -
\delta h \ \widetilde C \wedge \delta \hat k\right )\right ],
\eeq
where $\delta h=\delta G G^{-1},\ \delta h=diag\left(\delta x_i \zeta 
(z)\right)$. From the relation 
$\widetilde L\widetilde C=\widetilde C\hat k$ it follows that:
$$ 
{\rm Tr} \left( \widetilde C^{-1}[\delta h,\widetilde L ]\wedge\delta 
\widetilde C \right) = {\rm Tr} \left( \widetilde C^{-1} \delta h
\wedge(\widetilde L \delta \widetilde C-\delta \widetilde C\hat k) \right)= 
-{\rm Tr} \left( \widetilde C^{-1}\delta h\wedge(\delta \widetilde L 
\widetilde C-\widetilde C \delta \hat k) \right).  
$$ 
Therefore,
\beq 
\omega=\res_{z=0} {\rm Tr} \left(2\delta \widetilde L \wedge \delta h+ 
\widetilde C^{-1}\delta L\wedge \delta \widetilde C\right).  
\eeq 
The first term equals $\sum_i \delta x_i\wedge \delta p_i$.  
The last term equals 
\beq 
{1\over 2}\ {\rm Tr}\ C_0^{-1}\delta (B_i^+A_j)\wedge \delta C_0= 
{1\over 2}\  {\rm Tr}\left(C_0^{-1}(\delta B_j A_j)\wedge \delta C_0 + 
\delta C_0^{-1}\wedge(B_j^+ \delta A_j)C_0\right) , 
\label{80} \eeq 
where $C_0$ is the matrix $c_i^j$ of leading coefficients of the 
expansions (\ref{22}). From (\ref{nll}) it follows that 
$$\sum_{j=1}^N \left(A_{j\a}\delta c_j^k+\delta A_{j,\a}c_j^k\right)=0,\ \ \ 
\sum_{j=1}^N \left(c_{ki}^*\delta B_{j}^{\a}+
\delta c_{ki}^*B_{j}^{\a} \right)=0\ , $$ 
where $c_{jk}^*$ are matrix elements of $C_0^{-1}$. Substitution of the last 
formulae into (\ref{80}) completes the proof of the following theorem.  
\begin{th} The symplectic form $\omega$ given by (\ref{54}) equals 
\beq 
\omega=-\sum_{i=1}^N \left(\delta p_i\wedge \delta x_i+
\sum_{\a=1}^l \delta B_{i}^{\a}\wedge \delta A_{i\a}\right) .\label{90} \eeq 
\end{th} 
When the symplectic structure $\omega$ is identified with a standard, it can 
be directly checked that equations (\ref{scm}), (\ref{AB}) are Hamiltonian 
with respect to $\omega$ and with the Hamiltonian 
\beq 
H={1\over 2}\sum_{i=1}^N p_i^2-{1\over 2}\sum_{i\neq j} 
(b_i^+a_j)(b_j^+a_i)\wp(x_i-x_j). 
\eeq 
Nevertheless, we would like to show that the existence of a Hamiltonian 
for system (\ref{scm}), (\ref{AB}) can be proved in the framework of our 
approach without use of the explicit formula for the symplectic structure.  

By definition a vector field $\p_t$ on a symplectic 
manifold is Hamiltonian, if the contraction $i_{\p_t}\omega(X)=
\omega(X,\p_t)$ of the symplectic form 
is an exact one-form $dH(X)$. The function $H$ is the Hamiltonian 
corresponding to the vector field $\p_t$. The 
equations $\p_t L=[M,L]$, $\p_t k=0$, and the equation 
\beq 
\p_tC(t,P)=MC(t,P)+\mu(P,t)C(t,P) , \label{d}
\eeq 
where $\mu(P,t)$ is a scalar function, imply 
\beq 
i_{\p_t}\omega=\sum_{i=1}^N\res_{P_i} 
\left(<C^*(\delta L+\delta k(z))(M+\mu)C(z,k)>-<C^*(z,k) 
[M,L]\delta C> \right)dz,  \label{H} 
\eeq 
Note, that if a matrix $\Lambda(z)$ is holomorphic outside $P_j$, then 
the differential $<C^*\Lambda C >dz$ is holomorphic outside $P_j$, 
as well. 
Therefore, the sum of its residues at $P_j$ is equal to zero. Using that 
and the relations
$$
<C^* [M,L]\delta C>\ =\ <C^*M(L-k)\delta C>\ =\ 
<C^*(z,k) M(\delta k-\delta L)C>, 
$$
we get 
\beq 
i_{\p_t}\omega=\sum_{i=1}^N\res_{P_i} \left(<C^* (\delta L+\delta 
k)C>\mu(P,t)\right)dz=2\sum_{i=1}^N\res_{P_i} \delta k  \ \mu(P,t)dz. 
\eeq 
The singular part of the function $\mu(P,t)$ is equal to the singular
part of the eigenvalues of the second Lax operator 
\beq
\mu_j(z)=-z^{-2}+z^{-1}2(k_j(z))+O(1) , \label{71}
\eeq 
where $k_j(z)$ is the expansion of $k(z)$ at $P_j$ (see (4.8) in \cite{bab}).
Hence, 
\beq
i_{\p_t}\omega=2\ \res_{z=0}\  {\rm Tr}\left(z^{-2}\delta \hat k+z^{-1}\delta 
\hat k^2 \right)dz=2\ \delta {\rm Tr}\ \hat k^2=\ 2\delta\ {\rm Tr} \ L^2=
\delta H .  
\eeq

\medskip
Now, let us define a symplectic structure for spin generalization
of the elliptic RS system by the formula
\beq
\omega=\sum_{i=1}^N\res_{P_i} \left(<C^*(z,k) 
(\delta L(z) L^{-1}+\delta \ln k)\wedge \delta C(z)>\right)dz,  
\label{72}
\eeq
where $L$ and $C$ are the Lax operator (\ref{LaxL}) and its eigenvector.
\begin{th} The two-form (\ref{72}) equals
\beq
\omega=2\sum_{s=1}^{g+l-1}\delta z(\g_s)\wedge \delta \ln k(\g_s). \label{610}
\eeq
Equation (\ref{43})-(\ref{45}) are Hamiltonian with respect to this
symplectic structure with the Hamiltonian
\beq
H=\sum_{i=1}^N (b_i^+a_i).
\label{73}
\eeq
\end{th}
Note, that (\ref{610}) implies that $\omega$ is closed and does define a 
symplectic structure on ${\cal M}$. 
The proof of Theorem 3.3 goes almost identically to the previous case. 

At this moment,  we do not know for $l>1$ an explicit expression for 
$\omega$ in the original coordinates on ${\cal M}$. Formula (\ref{72})
contains the inverse matrix $L^{-1}$. For $l=1$ it can be written explicitly.
Namely, let $\hat L$ be the matrix
$$ \hat L_{ij}= f_i\  {\s(z+x_i-x_j)\over \s(x_i-x_j-\eta)}, $$
which is (up to gauge transformation (\ref{cal}) and a scalar factor) equal
to (\ref{LaxL}) for $l=1$. Then the entries of the inverse matrix 
equal:
$$
(\hat L^{-1})_{jm}=f_m^{-1}{\s(z+x_j-x_m-(N-2)\eta)\over
\s(z+\eta)\s(z-(N-1)\eta)}\ {\prod_{k\neq m}\s(x_j-x_k+\eta)
\prod_{k}\s(x_j-x_k-\eta)\over
\prod_{k\neq j}\s(x_j-x_k)\prod_{k\neq m}\s(x_m-x_k)}\ .
$$
This formula has been used for the proof of the following theorem.
\begin{th} Let $L$ be the matrix $L_{ij}=f_i\Phi(x_i-x_j-\eta,z)$,
where $\Phi$ is given by (\ref{1.5}). Then the form $\omega$ defined by
(\ref{72}) equals
\beq
\omega=\sum_i\delta \ln f_i\wedge \delta x_i+\sum_{i,j} V(x_i-x_j)
\delta x_i\wedge \delta x_j, \label{last}
\eeq
where $V(x)$ is defined in (\ref{42}). 
\end{th}
Our approach is evenly applicable to the nested Bethe ansatz equations.
It gives Hamiltonian version of the proof that discrete evolution
$x_i^n$ is a canonical transform of the RS symplectic structure {\ref{last}). 
This result was obtained for the first time in \cite {NRK} with the help
of Lagrangian interpretation of ({\ref{bet}).  

\medskip

\noindent
\bigskip{\Large \bf Acknowledgements}

\noindent
The author was partly supported by the INTAS grant 94-3006. 
The work is an updated version of a talk given on the workshop 
``Calogero-Moser-Sutherland models'' , CRM, Montreal, March 1997.

\end{document}